\documentclass[12pt]{article}
\usepackage{putex}
\usepackage{graphicx}

\begin{document}

\preprint{hep-th/0612143 \\ PUPT-2219}

\institution{PU}{Joseph Henry Laboratories, Princeton University, Princeton, NJ 08544}

\title{Momentum fluctuations of heavy quarks in the gauge-string duality}

\authors{Steven S. Gubser}

\abstract{Using the gauge-string duality, I compute two-point functions of the force acting on an external quark moving through a finite temperature bath of ${\cal N}=4$ super-Yang-Mills theory.  I comment on the possible relevance of the string theory calculations to heavy quarks propagating through a quark-gluon plasma.}

\date{January 2007}

\maketitle

\section{Introduction}
\label{INTRODUCTION}

Recent string theory computations \cite{Herzog:2006gh,Gubser:2006bz} of the drag force on a heavy quark, and a related calculation \cite{Casalderrey-Solana:2006rq} of momentum diffusion for non-relativistic heavy quarks, have raised the tantalizing prospect that the gauge-string duality might help us understand the dynamics of charm and bottom quarks propagating through the quark-gluon plasma (QGP) produced at the relativistic heavy ion collider (RHIC).  Earlier work in a somewhat similar spirit includes \cite{Sin:2004yx}.  Independently, a proposal was made in \cite{Liu:2006ug} for extracting the jet-quenching parameter $\hat{q}$ for ${\cal N}=4$ gauge theory from a Wilson loop calculation amenable to solution through techniques of classical string theory.\footnote{There exists some debate in the literature regarding the calculation of \cite{Liu:2006ug}.  Arguments from both sides of this debate can be found in \cite{Chernicoff:2006hi,LiuNew,Argyres:2006yz}.}  Subsequent work includes extensions to non-conformal theories \cite{Buchel:2006bv,Nakano:2006js,Talavera:2006tj,Gao:2006uf}, non-zero chemical potentials \cite{Herzog:2006se,Caceres:2006dj,Lin:2006au,Avramis:2006ip,Armesto:2006zv,Caceres:2006as}, and other deformations of ${\cal N}=4$ gauge theory \cite{Vazquez-Poritz:2006ba}; studies of directional emission \cite{Friess:2006aw,Friess:2006fk} and the relation to the magnetic string tension \cite{Sin:2006yz}; and calculations of drag on particles carrying higher representations of the gauge group \cite{Chernicoff:2006yp}.  The venue for the string theory calculations is the gauge-string duality \cite{Maldacena:1997re,Gubser:1998bc,Witten:1998qj} (for reviews see \cite{MAGOO,DHoker:2002aw,Klebanov:2000me}), in particular the computation of Wilson loops through dual classical configurations of fundamental strings in anti-de Sitter space, as first considered in \cite{Rey:1998ik,Maldacena:1998im}.

The drag force calculations \cite{Herzog:2006gh,Gubser:2006bz,Casalderrey-Solana:2006rq} result in a relaxation time $t_D$ for charm quarks that is in the right ballpark for comparison with data on the nuclear modification factor $R_{AA}$ and the elliptic flow parameter $v_2$ for heavy quarks, as reported most recently by the STAR and PHENIX collaborations in \cite{Abelev:2006db,Adare:2006nq}.  According to the prescription for comparing ${\cal N}=4$ gauge theory and QCD advocated in \cite{Gubser:2006qh}, the string theory prediction is $t_D = 2.1 \pm 1 \, {\rm fm}/c$ for charm quarks.  This value is lower than typical values $t_D \sim 4.5\,{\rm fm}/c$ used in one of the more successful phenomenological theories \cite{vanHees:2005wb} of heavy-quark transport.  But one should bear in mind that many steps separate the drag force calculation from predictions of $R_{AA}$ and $v_2$.  Prominent among these steps is a description of heavy quark propagation through the QGP which includes fluctuations in the force on the quark.  The goal of this paper is to calculate the two-point functions of these fluctuations using the gauge-string duality.

The organization of the rest of this paper is as follows.  Section~\ref{DIFFUSE} sets the stage by reviewing the diffusion of momentum of a heavy quark in a Langevin formalism.  Section~\ref{FLUCTUATE} sets up the main calculation by finding the quadratic action and equations of motion for fluctuations of the trailing string.  Section~\ref{CORRELATORS} shows how to translate appropriate solutions of these equations of motion into the Green's functions of interest.  Section~\ref{ZEROT} shows how to treat the zero-temperature case analytically.  Section~\ref{NUMERICS} presents a numerical study of the two-point function of primary interest.  Section~\ref{DISCUSSION} includes a discussion of how the string theory results might be applied to understanding heavy quarks propagating through a real-world quark-gluon plasma.  This discussion suffers from the usual difficulties of relating two significantly different theories, namely ${\cal N}=4$ super-Yang-Mills and QCD.  Some significant technical issues are postponed to three appendices.

\section{Momentum diffusion in a Langevin description}
\label{DIFFUSE}

Following \cite{Svetitsky:1987gq} one may attempt to describe the propagation of a heavy quark through a thermal bath in terms of Langevin dynamics (see also, for example, \cite{vanHees:2004gq,Moore:2004tg}):
 \eqn{Langevin}{
  {dp_i \over dt} &= -\eta_D(p) \, p_i + F_i^L + F_i^T  \cr
  \langle F_i^L(t_1) F_j^L(t_2) \rangle &= 
   \kappa_L(p) \, \hat{p}_i \hat{p}_j \delta(t_1-t_2)  \cr
  \langle F_i^T(t_1) F_j^T(t_2) \rangle &=
   \kappa_T(p) \, (\delta_{ij} - \hat{p}_i \hat{p}_j) 
     \delta(t_1-t_2) \,,
 }
where $\hat{p}_i = p_i/p$ is the unit 3-vector in the direction of the momentum $\vec{p}$.  The string theory prediction for the drag force (ignoring issues of modified dispersion relations raised in \cite{Herzog:2006gh}) amounts to
 \eqn{StringEta}{
  \eta_D = {\pi \over 2} \sqrt{g_{YM}^2 N}\, {T^2 \over m} \,,
 }
independent of $p$, where $m$ is the mass of the heavy quark.  It is often assumed (modulo a subtlety having to do with how one discretizes the Langevin equations in the process of passing to a Fokker-Planck description) that $\kappa_L$ and $\eta_D$ are constrained by the Einstein relation
 \eqn{EinsteinRelation}{
  \kappa_L = 2TE \, \eta_D = \pi \sqrt{g_{YM}^2 N}\, T^3 \gamma \,,
 }
where
 \eqn{gammaDef}{
  \gamma = {1 \over \sqrt{1-v^2}}
 }
is the standard Lorentz factor for the heavy quark (and I have again ignored the possibility of a modified dispersion relation).  The relation \eno{EinsteinRelation} emerges from requiring that a heavy quark propagating according to the equations \eno{Langevin} should eventually equilibrate to a thermal distribution $e^{-E/T}$, so it is a consistency condition for the Langevin approach.  No such relation is available for $\kappa_T$, but for non-relativistic quarks isotropy demands $\kappa_T = \kappa_L$.  Indeed, the result
 \eqn{kappaLT}{
  \kappa_L = \kappa_T = {2T^2 \over D} = 
    \pi \sqrt{g_{YM}^2 N}\, T^3 \qquad\hbox{for $v \ll 1$}
 }
was obtained through direct calculation in \cite{Casalderrey-Solana:2006rq} and through use of the Einstein relation in \cite{Herzog:2006gh}.
 
The Langevin approach \eno{Langevin} is hardly the only one in use for describing quark dynamics in the QGP.  Another prominent paradigm hinges on radiative energy loss \cite{Baier:1996kr,Baier:1996sk,Zakharov:1997uu}.  Central interest in this description attaches to the BDMPS jet-quenching parameter $\hat{q}$.  The jet-quenching parameter as I prefer to define it is
 \eqn{qhatDef}{
  \hat{q}_T = \langle p_\perp^2 \rangle / \lambda \,.
 }
Here $\langle p_\perp^2 \rangle$ is the average transverse momentum acquired by a parton after it has traveled a distance $\lambda$, measured in the rest frame of the plasma.\footnote{I would use $p_T$ in place of $p_\perp$ except for the fact that $p_{\rm T}$ is usually reserved for momentum perpendicular to the beampipe.  If the parton is traveling in the $x^1$ direction, then $\vec{p}_\perp$ is the projection of the momentum onto the $x^2$-$x^3$ plane.}  The path length $\lambda$ is supposed to be taken large enough so that short-range correlations are washed out, but short enough so that the quark is only slightly deflected from its original trajectory.  A definition of $\hat{q}$ that is preferred in some works, for example \cite{Liu:2006ug,LiuNew}, refers to a partially light-like Wilson loop.  To distinguish between the $\hat{q}$ of these works and the definition I prefer, I will employ a subscripted $T$ as in \eno{qhatDef}.  Note that the definition of $\hat{q}_T$ does not require a strict light-like limit: it can be evaluated at any $v$.  Also, does not require us to commit to the BDMPS formalism.  For example, consider a Langevin description of momentum broadening where a heavy quark travels initially in the $x^1$ direction and is acted on by stochastic transverse forces satisfying
 \eqn{PartonLangevin}{
  {dp_i \over dt} = F_i \qquad
   \langle F_i(t_1) F_j(t_2) \rangle = \delta_{ij} K_T(t_1-t_2) \,,
 }
where $K_T(t)$ is an integrable function and $i,j$ run over values $2,3$.  If one assumes $p_\perp(0) = 0$, then at sufficiently late times $t$ (meaning times larger than the characteristic time-scales of $K_T(t)$) one has
 \eqn{MomentumGrowth}{
  \langle p_\perp(t)^2 \rangle 
    = \delta^{ij} \langle p_i(t) p_j(t) \rangle
    = \int_0^t dt_1 \int_0^t dt_2 \, 2 K_T(t_1-t_2)
    \approx 2\kappa_T t \,,
 }
where by definition
 \eqn{kappaDef}{
  \kappa_T = \int_{-\infty}^\infty dt \, K_T(t) \,.
 }
Comparing with \eno{qhatDef}, one extracts
 \eqn{qhatExtracted}{
  \hat{q}_T = {2\kappa_T \over v} \,.
 }
For a heavy quark with a known initial momentum, one may also consider
 \eqn{qhatLdef}{
  \hat{q}_L = \langle (\Delta p_L)^2 \rangle / \lambda
 }
where $\Delta p_L$ is the deviation of the longitudinal momentum of a heavy quark from some average value.  As with $\hat{q}_T$, the definition \eno{qhatLdef} doesn't commit us to a specific formalism; however, a Langevin analysis precisely analogous to the one leading to \eno{qhatExtracted} gives
 \eqn{qhatLextracted}{
  \hat{q}_L = {\kappa_L \over v}
 }
where
 \eqn{kappaLDef}{
  \kappa_L = \int_{-\infty}^\infty dt \, \langle F_1(t) F_1(0)
    \rangle
 }
and $F_1$ is the fluctuating part of the longitudinal force.

The AdS/CFT calculations that are the focus of sections~\ref{FLUCTUATE}-\ref{NUMERICS} lead to force correlators from which $\hat{q}_L$ and $\hat{q}_T$ may be extracted, essentially as in \eno{kappaDef}, \eno{qhatExtracted}, \eno{qhatLextracted}, and \eno{kappaLDef}.  The results are
 \eqn{HugeKappaL}{
  \hat{q}_T = 2\pi \sqrt{g_{YM}^2 N}\, T^3 \sqrt\gamma/v \qquad
  \hat{q}_L = \pi \sqrt{g_{YM}^2 N}\, T^3 \gamma^{5/2}/v \,.
 }
The result \eno{HugeKappaL} is larger by a factor $1/(1-v^2)^{3/4}$ than expected from \eno{EinsteinRelation}.  So Langevin dynamics does not capture the whole story.  The method for obtaining \eno{HugeKappaL} is to numerically compute a symmetrized Wightman two-point function of oscillations of the trailing string solution of \cite{Herzog:2006gh,Gubser:2006bz} and identify the zero-frequency component of this two-point function with $\kappa_L$ or $\kappa_T$ as appearing in \eno{kappaDef} and \eno{kappaLDef}.  The expressions in \eno{HugeKappaL} are obtained as a fit to the low-frequency limit of the numerics, but they are probably exact statements about the Green's functions in question.\footnote{I thank C.~Herzog for first suggesting to me the comparison with the non-relativistic results \eno{kappaLT}.}

\section{Fluctuations of the trailing string}
\label{FLUCTUATE}

The drag force prediction from the gauge-string duality,
 \eqn{DragForce}{
  {d\vec{p} \over dt} = -\eta_D \vec{p} = 
    -{\pi \over 2} \sqrt{g_{YM}^2 N}\, T^2 {\vec{p} \over m} \,,
 }
arises from a classical string configuration \cite{Herzog:2006gh,Gubser:2006bz} in the $AdS_5$-Schwarzschild background,
 \eqn{AdSSchwarzschild}{
  ds^2 = {L^2 \over z_H^2 y} \left( -h dt^2 + d\vec{x}^2 + 
    {z_H^2 dy^2 \over h} \right) \qquad
  h \equiv 1-y^4 \qquad z_H = {1 \over \pi T} \,,
 }
where $T$ is the Hawking temperature.  The string configuration, for $\vec{p}$ pointing in the $x^1$ direction, is described by
 \eqn{TrailingString}{
  x^1 = x_0^1(t,y) \equiv vt + {v z_H \over 2} \left( \tan^{-1} y + 
      \log \sqrt{1-y \over 1+y} \right) \,.
 }
The fluctuating part of the force on the quark, which is located at the $y=0$ end of the string (i.e.~on the boundary of $AdS_5$-Schwarzschild) should be computable in terms of linearized fluctuations around the solution \eno{TrailingString}.  More specifically, if
 \eqn{Fluctuated}{
  x^1 = x_0^1(t,y)+\delta x^1(t,y) \qquad
   x^2 = \delta x^2(t,y) \qquad
   x^3 = \delta x^3(t,y) \,,
 }
then the Nambu-Goto action,
 \eqn{NGaction}{
  S_{\rm NG} = -{1 \over 2\pi\alpha'} \int d^2\sigma \, 
    \sqrt{-\det g_{\alpha\beta}} \qquad
    g_{\alpha\beta} = G_{\mu\nu} \partial_\alpha x^\mu 
      \partial_\beta x^\nu \,,
 }
may be expanded to quadratic order in $\delta x^i$ to obtain
 \eqn{NGexpanded}{
  S_{\rm NG} &= -{L^2 \over 2\pi\alpha'} {1 \over z_H} 
    \int dt dy \, {y_S^2 \over y^2} + 
    \int dt dy \, {\cal P}^\alpha \partial_\alpha \delta x^1  \cr
   &\qquad{} - {1 \over 2} \int dt dy \, \left( 
     {\cal G}_L^{\alpha\beta} \partial_\alpha \delta x^1 
       \partial_\beta \delta x^1 + 
     \sum_{i=2,3} {\cal G}_T^{\alpha\beta} \partial_\alpha \delta x^i
       \partial_\beta \delta x^i \right) \,,
 }
where
 \eqn{Pdef}{
  {\cal P}^\alpha = {L^2/z_H^2 \over 2\pi\alpha'} {v \over y_S^2}
    \begin{pmatrix} z_H / y^2 (1-y^4) \\ 1 \end{pmatrix} \,,
 }
 \eqn{Gdef}{
  {\cal G}_T^{\alpha\beta} = y_S^4 
    {\cal G}_L^{\alpha\beta} = 
    {L^2/z_H^2 \over 2\pi\alpha'} {1 \over y_S^2}
    \begin{pmatrix} \displaystyle{{z_H \over y^2} 
       {1-y^4 y_S^4 \over (1-y^4)^2}} & 
      \displaystyle{v^2 \over 1-y^4}  \\[4\jot]
     \displaystyle{v^2 \over 1-y^4} & 
      \displaystyle{y^4-y_S^4 \over y^2 z_H}
    \end{pmatrix} \,,
 }
and I have introduced
 \eqn{ySdef}{
  y_S \equiv \sqrt[4]{1-v^2} \,.
 }
If I switch from $\sigma^\alpha = (t,y)$ to some other parametrization of the worldsheet, then ${\cal P}^\alpha$, ${\cal G}_L^{\alpha\beta}$, and ${\cal G}_T^{\alpha\beta}$ would transform not as tensors but as tensor densities: that is, they include a factor of $\sqrt{-\det g_{\alpha\beta}}$.

Because $\partial_\alpha {\cal P}^\alpha = 0$ and ${\cal G}_L^{\alpha\beta}$ is proportional to ${\cal G}_T^{\alpha\beta}$, all three $\delta x^i(t,y)$ obey the same equation of motion, namely 
 \eqn{phiEOM}{
  \partial_\alpha ({\cal G}_T^{\alpha\beta} \phi) = 0
 }
where $\phi = \delta x^i$, $i=1,2,3$.  Plugging an expansion
 \eqn{SeparateAnsatz}{
  \phi(t,y) = \int_{-\infty}^\infty {d\omega \over 2\pi} 
    \phi_0(\omega) \Psi(\omega,t,y) \qquad
    \Psi(\omega,t,y) = e^{-i\omega t} \psi(\omega,y)
 }
into \eno{phiEOM} leads to a ``radial'' equation for $\psi$:
\eqn{psiEOM}{
  \left[ s(y) \partial_y^2 + t(\omega,y) \partial_y + 
    u(\omega,y) \right] \psi(\omega,y) = 0 \,,
 }
where
 \eqn{stuDef}{
  s(y) &= -y (1-y^4)^2 (y_S^4-y^4)  \cr
  t(\omega,y) &= 2 (1-y^4) \left[ 1 - y^8 - 
     v^2 (1-y^4+iy^3 \omega z_H) \right]  \cr
  u(\omega,y) &= -y \omega z_H \left[ (1-y^4) \omega z_H + 
    v^2 y^4 (4iy + \omega z_H) \right] \,.
 }
The radial equation \eno{psiEOM} has regular singular points at the zeroes of $s(y)$: $y=0$, $y=\varkappa$, and $y=\varkappa y_S$ where $\varkappa$ is any fourth root of unity.  The most interesting of these is $y=y_S$, corresponding to some intermediate point on the string.  What is special about this point is that it is the location of a horizon of the induced metric on the worldsheet.  An intuitive way to see this is that a point on the trailing string with $y$ held fixed follows a timelike trajectory if $y<y_S$ and a spacelike one if $y>y_S$.\footnote{A closely related observation \cite{Herzog:2006gh,Peeters:2006iu,Liu:2006nn,Chernicoff:2006hi} is that there are no-drag configurations of mesons represented as strings with both ends attached to branes in $AdS_5$-Schwarzschild provided the string between the quarks hangs no lower than $y=y_S$.}  So the region $y<y_S$ corresponds to the exterior of the ``black hole on the worldsheet,'' and $y>y_S$ corresponds to the interior.  No signal from the interior can propagate classically to the exterior.  Instead, signals in the interior region must by causality travel down the string.  But there should be some Hawking radiation from the worldsheet horizon upward toward $y=0$, and it is natural to suppose that it relates to the momentum diffusion.  I will not make this connection directly, but the two-point functions that I will compute are related to both classical absorption and spontaneous emission by the worldsheet horizon.

Although \eno{psiEOM} does not seem to be explicitly solvable, certain limits of it are tractable.  For example, to next-to-leading order in small $y$, two independent solutions are
 \eqn{FarSoln}{
  \psi_{\rm Def}(\omega,y) = 1 + {\omega^2 z_H^2 \over 2 y_S^4} 
    y^2 + O(y^4) \qquad
  \psi_{\rm VEV}(\omega,y) = y^3 - {\omega^2 z_H^2 \over 
    10 y_S^4} y^5 + O(y^6) \,,
 }
and to leading order in small positive $y_S-y$, two independent solutions are
 \eqn{NearSoln}{
  \psi_-(\omega,y) = 1 + O(y_S-y) \qquad
  \psi_+(\omega,y) = (y_S-y)^{i\omega z_H/2y_S} \left[ 1 + 
    O(y_S-y) \right] \,.
 }
Of the two solutions in \eno{NearSoln}, $\psi_+(\omega,y)$ corresponds to an outgoing wave, because the phase increases as one goes to smaller values of $y$.  The standard horizon boundary condition, then, is to disallow this solution.  To justify this statement completely, one should show that suppressing this solution amounts to stipulating that $\Psi(\omega,t,y)$ should depend only on the infalling coordinate at the horizon.  This is indeed true, because, to leading order close to the horizon, the infalling and outgoing coordinates are
 \eqn{InOutCoordinates}{
  u_- = t \qquad u_+ = t - {z_H \over 2 y_S} \log(y_S-y) \,.
 }
A justification of \eno{InOutCoordinates} is postponed to appendix~\ref{WORLDSHEET}.

Consider now a set of wave-functions $\Psi_R(\omega,t,y) = e^{-i\omega t} \psi_R(\omega,y)$, defined so that
 \eqn{psiBCbdy}{
  \psi_R(\omega,y) = \psi_{\rm Def}(\omega,y) + 
    C_R(\omega) \psi_{\rm VEV}(\omega,y) = 
    C^H_-(\omega) \psi_-(\omega,y) \,.
 }
In \eno{psiBCbdy}, $\psi_{\rm Def}$, $\psi_{\rm VEV}$, and $\psi_\pm$ are regarded as exact solutions of the radial equation \eno{psiEOM}, specified by their asymptotics at $y=0$ or $y=y_S$.  In the absence of an analytical solution to \eno{stuDef}, the quantities $C_R(\omega)$ and $C^H_-(\omega)$ must be determined through some approximation scheme or through numerics.  The wave-functions $\Psi_R(\omega,t,y)$ have several useful properties:
 \begin{enumerate}
  \item $\Psi_R(\omega,t,y)$ is a solution of the wave equation \eno{phiEOM} for $\phi$.
  \item $\Psi_R(\omega,t,0) = e^{-i\omega t}$.
  \item $\Psi_R(\omega,t,y)^*=\Psi_R(-\omega,t,y)$.  To see this, note that complex conjugation is equivalent to sending $\omega \to -\omega$ in the radial equation \eno{psiEOM}, and that the boundary condition at the horizon is preserved by complex conjugation.  This property guarantees that $\phi(t,y)$ is real everywhere provided $\phi_0(\omega)^* = \phi_0(-\omega)$.\label{RealityPsi}
 \end{enumerate}

\section{Momentum correlators from the trailing string}
\label{CORRELATORS}

As in the previous section, let $\phi$ be one of the $\delta x^i$.  Let ${\cal O}$ be the operator dual to $\phi$, and let ${\cal G}^{\alpha\beta} = {\cal G}_L^{\alpha\beta}$ or ${\cal G}_T^{\alpha\beta}$, as appropriate.  The retarded, time-ordered, and symmetrized Wightman two-point functions are defined as
 \eqn[c]{SeveralGreen}{
  G^R(t) = -i \theta(t) \langle [{\cal O}(t),{\cal O}(0)] \rangle
    \qquad
  G^F(t) = -i \langle T \{ {\cal O}(t) {\cal O}(0) \} \rangle  \cr
  G(t) = {1 \over 2} \langle {\cal O}(t) {\cal O}(0) + 
    {\cal O}(0) {\cal O}(t) \rangle
 }
where in general
 \eqn{averagesDefined}{
  \langle Q \rangle = {\tr e^{-\beta H} Q \over \tr e^{-\beta H}} \,,
 }
and $H$ is the Hamiltonian of the gauge theory.

In analogy to the recipe of \cite{Son:2002sd}, the proposal for extracting the retarded Green's function from the wave-functions $\Psi_R(\omega,t,y)$ is
 \eqn{GRextract}{
  G^R(\omega) = -\Psi_R^*(\omega,t,y) {\cal G}^{y\beta}
    \partial_\beta \Psi_R(\omega,t,y) \Big|_{y=0} = 
   -{\cal G}^{y\beta} \partial_\beta \log
     \Psi_R(\omega,t,y) \Big|_{y=0} \,,
 }
where the notation $|_{y=0}$ means to evaluate at $y=0$ after taking the derivatives.  The difference between \eno{GRextract} and (3.15) of \cite{Son:2002sd} is the use of $\Psi_R$ rather than $\psi_R$.\footnote{Actually there seems to be one further difference: an overall sign.  Possibly I have misunderstood the notation used \cite{Son:2002sd}.  Anyway, I claim that the sign in \eno{GRextract} is the right one.}  Using $\psi_R$ would amount to restricting the sum over $\beta$ to $\beta=y$, and this does not make sense in light of worldsheet reparametrization invariance.  If the number current associated with $\Psi_R(\omega,t,y)$ is defined as
 \eqn{NumberPsi}{
  J^\alpha = {1 \over 2i} {\cal G}^{\alpha\beta} \Psi_R^*
    \overleftrightarrow\partial\!\!{}_\beta \Psi_R \,,
 }
then
 \eqn{ImGR}{
  \Im G^R(\omega) = -J^y \,,
 }
and because $J^\alpha$ is conserved, the right hand side of \eno{ImGR} can be evaluated at any $y$.  The number flux is positive at the horizon for $\omega>0$ as a consequence of the infalling boundary conditions, so one finds from \eno{ImGR} that
 \eqn{ImGRzero}{
  \Im G^R(\omega) < 0 \qquad\hbox{for $\omega>0$.}
 }
This is the correct sign for describing dissipative dynamics.  If $\psi_R$ had been used instead of $\Psi_R$, the connection with the conserved number current would be broken.  In \cite{Son:2002sd} a restriction was made to consider only diagonal metrics, so the question of including dependence of the wave-function on coordinates other than the radial one never arose.  (The context was slightly different: instead of ${\cal G}^{\alpha\beta}$, the metric of interest was the line element of the bulk spacetime.)  But \eno{GRextract} is clearly in the spirit of \cite{Son:2002sd} and its antecedent \cite{Gubser:1998bc}: note for instance the similarity with  (27) of \cite{Gubser:1998bc}.

The quadratic part of the on-shell action reduces to
 \eqn{OnshellAction}{
  S_{\rm on-shell} = {1 \over 2} \int \phi_0(-\omega) \phi_0(\omega)
    \Re G^R(\omega) \,,
 }
where to obtain the right hand side from the quadratic part of $S_{\rm NG}$ I used an integration by parts, the equations of motion, the formula \eno{GRextract}, and the property $G^R(-\omega) = G^R(\omega)^*$, which is obvious from property \ref{RealityPsi} at the end of section~\ref{FLUCTUATE}.

Thus, just as in the examples treated in \cite{Son:2002sd}, the real part of $G^R(\omega)$ may be extracted using a conventional AdS/CFT formulation in which the on-shell action is the generating functional for Green's functions.  But it turns out that the imaginary part will be of greater interest, due to its connection with the dissipative dynamics.  The relation between $\Im G^R(\omega)$ and $G(\omega)$,
 \eqn{WightmanFromGR}{
  G(\omega) = -\coth {\omega \over 2 T y_S} \Im G^R(\omega) \,,
 }
is modified from the usual one in that $T y_S$ appears in place of $T$.  A justification of \eno{WightmanFromGR}, following \cite{Casalderrey-Solana:2007qw}, is postponed to Appendix~\ref{SCHWINGER}.

Here is a somewhat informal argument for identifying ${\cal O}(t)$ with minus the force on the quark in the direction that $\phi$ selects.  To be precise, let $\phi=\delta x^2$: then the claim is ${\cal O}(t) = -\dot{p}_2(t)$.  Consider the action
 \eqn{Squark}{
  S_q[\phi(t)] = \int dt \, L(\dot\phi_0)
 }
for an external quark propagating through the thermal medium along a path
 \eqn{QuarkPath}{
  x^1(t) = vt \qquad x^2(t) = \phi_0(t) \,.
 }
I have excluded explicit dependence on $\phi_0$ and $t$ from $L$ because the medium is assumed to be translationally invariant and static.  Note that translational invariance does not imply conservation of transverse momentum: $S_q$ is evaluated under the path integral and includes couplings to the gauge theory.  For small deviations $\phi_0$ from a straight path,
 \eqn{SquarkApprox}{
  S_q \approx \int dt \, \dot\phi_0
     {\partial L \over \partial\dot\phi_0}
   = \int dt \, \dot\phi_0 p_\phi
   = -\int dt \, \phi_0 \dot{p}_\phi \,.
 }
Of course, $p_\phi = p_2$.  Altogether, the generating functional for the two-point functions of interest is
 \eqn{GeneratingZ}{
  Z[\phi_0] = \left\langle \exp\left\{ 
    -i \int dt \, \phi_0 \dot{p}_2 \right\} \right\rangle \,,
 }
whence the desired conclusion ${\cal O} = -\dot{p}_2$.  More formally, $p_2(t)$ should be regarded as the operator which generates a local displacement at time $t$ of the path of the Wilson loop for the external quark.  A similar line of argument applies when $\phi = \delta x^1$, and of course $\delta x^3$ is equivalent to $\delta x^2$.

Other Green's functions than $G^R(\omega)$ could be computed by using other wave-functions than $\Psi_R(\omega,t,y)$ in \eno{GRextract}, provided they satisfy the three properties at the end of section~\ref{FLUCTUATE}.  For example, wave-functions $\Psi_A(\omega,t,y) = e^{-i\omega t} \psi_A(\omega,y)$ with $\psi_A$ proportional to $\psi_+$ should lead to the advanced Green's function.

\section{Zero-temperature Green's functions}
\label{ZEROT}

As a check on the validity of the prescription \eno{GRextract}, and as a way of understand the $\omega \gg T$ behavior in the $T \neq 0$ case, it is useful to detour to a consideration of the zero temperature limit, where the background metric is just
 \eqn{PureAdS}{
  ds^2 = {L^2 \over z^2} (-dt^2 + d\vec{x}^2 + dz^2)
 }
and the unperturbed solution is a string that hangs straight down:
 \eqn{StraightDown}{
  x^1 = vt \qquad x^2 = x^3 = 0 \,.
 }
I will use worldsheet coordinates $\sigma^\alpha = (t,z)$.  An expansion entirely analogous to \eno{NGexpanded} can be performed, only now the integration measure is $dtdz$, and 
 \eqn{PGG}{
  {\cal P}^\alpha = 0 \qquad
  {\cal G}_T^{\alpha\beta} = (1-v^2) {\cal G}_L^{\alpha\beta} = 
    {L^2/z^2 \over 2\pi\alpha'} \gamma 
     \begin{pmatrix} 1 & 0 \\ 0 & -1/\gamma^2 \end{pmatrix} \,.
 }
The wave-functions satisfying infalling boundary conditions at the degenerate horizon at $z=\infty$ were found analytically for $v=0$ in \cite{Callan:1999ki} and can be adapted immediately to the case of interest:
 \eqn{FoundPsiR}{
  \Psi_R(\omega,t,z) = e^{-i\omega t} e^{i\gamma\omega z/y_S^2}
    \left( 1 - i\gamma\omega z \right) \,.
 }
The retarded Green's functions are
 \eqn{GRzeroT}{
  G_T^R(\omega) = {1 \over \gamma^2} G_L^R(\omega) = 
    -{iL^2 \gamma^2 \over 2\pi\alpha'} \omega^3 \,,
 }
where I have dropped a divergent term proportional to $\omega^2$ whose Fourier transform is a contact term supported at $t=0$.  The associated symmetrized Wightman functions are
 \eqn{GWzeroT}{
  G_T(\omega) = -(\sgn\omega) \Im G_T^R(\omega) = 
    {L^2 \gamma^2 \over 2\pi\alpha'} |\omega|^3 \qquad
  G_L(\omega) = {L^2 \gamma^4 \over 2\pi\alpha'} |\omega|^3 \,.
 }
In preparation for finding the real-time forms $G_T(t)$ and $G_L(t)$, consider the general power-law Fourier integral:
 \eqn{ToExpand}{
  \int_{-\infty}^\infty dt \, {e^{i\omega t} \over 
      |t|^{2\Delta}}
    = \int_{-\infty}^\infty dt \, {2\cos \omega t \over 
        |t|^{2\Delta}}
    = 2 |\omega|^{2\Delta-1} \, \Gamma(1-2\Delta) \sin\pi\Delta \,.
 }
Expanding around $\Delta=2$ gives 
 \eqn{ExpandedInt}{
  \int_{-\infty}^\infty dt \, {e^{i\omega t} \over t^4} = 
    {\pi \over 6} |\omega|^3 + \hbox{(analytic in $\omega$)} \,.
 }
There is an analytic continuation implicit in the result \eno{ExpandedInt}: whereas the integral \eno{ToExpand} can be evaluated for $0<\Delta<1/2$ with no regularization, the integral in \eno{ExpandedInt} is highly divergent.  Regularization schemes differ in how they prescribe the analytic terms on the right hand side of \eno{ExpandedInt}, corresponding to contact terms in real time, supported at $t=0$.  Employing \eno{ExpandedInt} and ignoring contact terms, one obtains
 \eqn{GTGLtime}{
  G_T(t) = {1 \over \gamma^2} G_L(t) = 
    {3L^2\gamma^2 \over \pi^2 \alpha'} {1 \over t^4} \,.
 }

To me the most surprising feature of the final result \eno{HugeKappaL} in the case of non-zero temperature is the strong dependence of the magnitude of longitudinal fluctuations on $\gamma$.  One can see already in \eno{GRzeroT} and \eno{GTGLtime} a very strong dependence: $G_L^R \propto \gamma^4$ for fixed $\omega$ or $t$.  In this zero-temperature setting, the factors of $\gamma$ should be capable of being understood just in terms of Lorentz invariance.  As a step in this direction, consider the following toy model: ${\cal N}=4$ super-Yang-Mills with gauge group $U(1)$.  The action for the heavy quark is
 \eqn{SheavyQuark}{
  S = \int ds \, \left[ -m |\dot{x}| + A_\mu \dot{x}^\mu + 
   |\dot{x}| \theta^I X^I \right] \,,
 }
where dots denote $d/ds$ and the parameter $s$ need not be the proper time $\tau$.  The $\theta^I$ form a unit vector in ${\bf R}^6$, and they specify properties of the quark: its scalar ``charges.''  For an external quark, one specifies the $\theta^I$ as functions of $s$ along with the trajectory $x^\mu(s)$.  The only case I will consider (which is also the case considered in \cite{Herzog:2006gh,Casalderrey-Solana:2006rq,Gubser:2006bz}) is the one where the $\theta^I$ are constant.  The classical equation of motion following from \eno{SheavyQuark} is
 \eqn{QuarkEOM}{
  m \dot{u}_\mu = {\cal W}_\mu + \theta^I X^I \dot{u}_\mu \,,
 }
where I have now specialized to the case where $s$ is the proper time $\tau$, and
 \eqn{Wdef}{
  {\cal W}_\mu \equiv F_{\mu\nu} u^\nu + \theta^I {\partial X^I \over
    \partial x^\nu} (\delta^\nu_\mu + u^\nu u_\mu) \,.
 }
The velocity vector $u^\mu = dx^\mu/d\tau$ satisfies
 \eqn{uSquared}{
  \eta_{\mu\nu} u^\mu u^\nu = -1 \qquad 
   \eta_{\mu\nu} = \diag\{ -1,1,1,1 \} \,.
 }
Let's now promote $F_{\mu\nu}(t,\vec{x})$ and $X^I(t,\vec{x})$ to operator-valued fields and consider the symmetrized Wightman two-point function of ${\cal W}_\mu(\tau)$.  Because of conformal invariance and the obvious identity ${\cal W}_\mu u^\mu = 0$, the answer can only be
 \eqn{Wcorrelator}{
  G^{\cal W}_{\mu\nu}(\tau) \equiv {1 \over 2} 
   \langle {\cal W}_\mu(\tau) {\cal W}_\nu(0) + 
     {\cal W}_\nu(0) {\cal W}_\mu(\tau) \rangle = 
    C_{\cal W} {\eta_{\mu\nu} + u_\mu u_\nu \over \tau^4}
 }
for some constant $C_{\cal W}$.

Whether ${\cal W}_\mu$ is the entire ``force'' acting on the quark depends on whether the last term in \eno{QuarkEOM} is considered a force or a modification of the momentum.  This is immaterial for external quarks because their mass is infinite and their trajectory can be prescribed to have constant $u^\mu$.  So for the purposes of comparing \eno{Wcorrelator} to \eno{GTGLtime}, this term doesn't matter.  What does matter is that, in deriving \eno{GTGLtime}, the force was defined in terms of a $t$ derivative, not a $\tau$ derivative:
 \eqn{ForceDefine}{
  F_i = {dp_i \over dt} = {1 \over \gamma} {dp_i \over d\tau} = 
    \gamma {\cal W}_i \,.
 }
Using \eno{Wcorrelator} and~\eno{ForceDefine}, and assuming as usual that the quark moves in the $x^1$ direction, one obtains
 \eqn{Fonetwo}{
  G_L(t) = {1 \over 2} \langle F_1(t) F_1(0) + F_1(0) F_1(t) \rangle
     = {1 \over \gamma^2} G^{\cal W}_{11}(t/\gamma)
     = \gamma^4 {C_{\cal W} \over t^4} \,,
 }
and similarly $G_T(t) = \gamma^2 C_{\cal W}/t^4$.  Thus there is perfect agreement with \eno{GTGLtime} if one formally makes the replacement $C_{\cal W} \to 3L^2/\pi^2\alpha'$.

The simple-minded analysis leading to \eno{Fonetwo} of course makes no prediction about the scaling of the low-frequency parts of $G_L(\omega)$ and $G_T(\omega)$ with $\gamma$ when the temperature is non-zero.  But what drives the relation $G_L(\omega) = \gamma^2 G_T(\omega)$ is the corresponding relation ${\cal G}^{\alpha\beta}_L = \gamma^2 {\cal G}^{\alpha\beta}_T$, and this holds whether $T$ is zero or not.

\section{Numerical results}
\label{NUMERICS}

\begin{figure}
  \centerline{\includegraphics[width=6.5in]{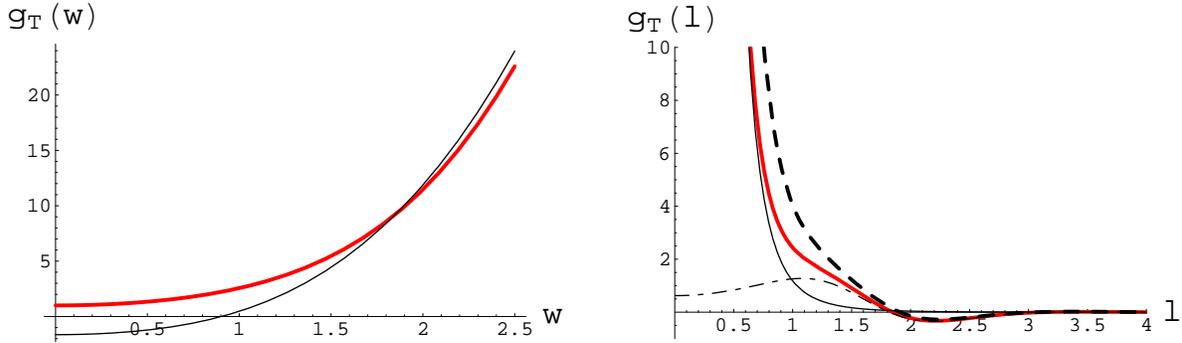}}
  \caption{The thick red lines come from numerical evaluations of $g_T(w)$ and the most accurate form of $g_T(\ell)$ I was able to obtain.  The thin black lines are the analytic approximations $g^{\rm approx}_T(w)$ and $g^{\rm approx}_T(\ell)$.  The thin dash-dot line in the $g_T(\ell)$ plot is the residual, $g_T(\ell)-g^{\rm approx}_T(\ell)$.  The thick dashed line is an approximation to $g_T(\ell)$ obtained by subtracting off only the leading $|w|^3$ behavior from $g_T(w)$ and numerically Fourier transforming the remainder out to $W=20$.}\label{gtSummary}
 \end{figure}

Starting from \eno{Gdef}, \eno{psiBCbdy}, \eno{GRextract}, and \eno{WightmanFromGR}, a short calculation shows that
 \eqn{GotG}{
  G_T^R(\omega) = -Y (2C_R(\omega) + iX) \qquad
   G_T(\omega) = Y \coth {\omega \over 2T y_S} \,
     (2\Im C_R(\omega) + X)
 }
where
 \eqn{XYdef}{
  X \equiv {2 \over 3} v^2 \gamma^2 \omega z_H \qquad
  Y \equiv {3 \over 4\pi\gamma z_H^3} {L^2 \over \alpha'}
    = {3 \over 4\pi\gamma z_H^3} \sqrt{g_{YM}^2 N}\,.
 }
The longitudinal Green's functions are always $\gamma^2$ times the transverse ones.  Only $G_T(\omega)$ will be evaluated explicitly here.  To do so, it helps to note that the substitutions
 \eqn[c]{chiYdefs}{
  \psi(\omega,y) = \left( {1+y \over 1-y} e^{-2\tan^{-1} y} 
    \right)^{i\omega z_H/4} \chi(w,Y)  \cr\noalign{\vskip2\jot}
  Y = {y \over y_S} \qquad
    w = {\omega z_H \over y_S}
 }
lead to a simplified form of the equation of motion \eno{psiEOM}:
 \eqn{SimplerEOM}{
  \left[ Y(1-Y^4) \partial_Y^2 - 2 (1+Y^4-iwY^3) \partial_Y + 
    w^2 Y \right] \chi(w,Y) = 0 \,.
 }
An infalling solution $\psi_R(\omega,y)$, satisfying the asymptotics \eno{psiBCbdy}, corresponds to a solution $\chi_R(w,Y)$ satisfying
 \eqn{chiAsymptotics}{
  \chi_R(w,Y) &= 1 + {w^2 Y^2 \over 2} + c_R(w) Y^3 + 
    O(Y^4) \qquad\hbox{for small $Y$}  \cr
  \chi_R(w,Y) &= c^H_-(w) \left[ 1 + O(1-Y) \right] 
    \qquad\hbox{for $Y$ close to $1$.}
 }
From \eno{GotG}, \eno{chiYdefs}, and \eno{chiAsymptotics} it is straightforward to show that
 \eqn{BestFormGTR}{
  G_T^R(\omega) = -{3 \over 4\pi z_H^3} 
    {\sqrt{g_{YM}^2 N} \over y_S} \left( 2 c_R(w) + 
     {2i \over 3} w \right)
 }
and that
 \eqn{BestFormG}{
  G_T(\omega) = {1 \over \pi^2 z_H^3}
    {\sqrt{g_{YM}^2 N} \over y_S} g_T(w)
 }
where
 \eqn{gTdef}{
  g_T(w) = \left( {\pi w \over 2} \coth {\pi w \over 2} \right)
   \left( 1 + {3 \Im c_R(w) \over w} \right) \,.
 }
To calculate $c_R(w)$, and thereby $g_T(w)$, I employed numerical integration, implemented with Mathematica's {\tt NDSolve}, matched to high-order series expansions for $Y$ close to $0$ and $1$.  Asymptotic forms at large and small $w$ were extracted by fitting to the numerics:
 \eqn{gtForms}{
  g_T(w) &= 1 + 1.26 w^2 + O(w^4) \qquad\hbox{for small $w$}  \cr
  g_T(w) &= {\pi \over 2} |w|^3 \left(1 - {1 \over w^4} + 
    O(w^{-8}) \right) 
    \qquad\hbox{for large $w$} \,.
 }
Graphs of $g_T(w)$ and its Fourier transform,
 \eqn{gtReal}{
  g_T(\ell) \equiv \int_{-\infty}^\infty {dw \over 2\pi} 
    e^{-iw\ell} g_T(w) \,,
 }
are shown in figure~\ref{gtSummary}.  By visual inspection, the scale $\ell_*$ for the characteristic structures in $g_T(\ell)$ is $\ell_* \approx 1.5$.

Because of the leading $|w|^3$ behavior in $g_T(w)$ at large $w$, the real space form $g_T(\ell)$ is highly divergent at small $\ell$.  This leading divergence is precisely the behavior observed in the zero-temperature Green's function in section~\ref{ZEROT}.  Contact terms supported at $\ell=0$ restore integrability at the origin.  A simple strategy for finding $g_T(\ell)$ is to subtract the $|w|^3$ term from $g_T(w)$ and perform a numerical Fourier transform on the remainder, cut off at some maximum frequency $W$.  The trouble with this strategy is that the subleading $1/|w|$ behavior in $g_T(w)$ leads to errors of order $\log (W|\ell|)$ at small $\ell$.  Subtracting this subleading behavior as well leads to a remainder which is absolutely integrable at large $w$ but not at small $w$.  The simplest form I could find that agrees with the large $w$ asymptotics in \eno{gtForms} to the order shown and leads to an everywhere absolutely integrable remainder is
 \eqn{FancyApprox}{
  g^{\rm approx}_T(w) &= 
    {\pi \over 2} |w|^3 \sqrt{1 + {1 \over \ell_0^2 w^2}}
    \left( 1 - {1 \over 2\ell_0^2 w^2} \right) 
      \cr
  g^{\rm approx}_T(\ell) &= {3 \over \ell^4} e^{-|\ell|/\ell_0}
    \left( 1 + {\ell^2 \over 4\ell_0^2} \right) 
    \qquad\hbox{where $\ell_0 = (3/8)^{1/4}$.}
 }
The intuition of Debye screening suggests that $g_T(\ell)$ decays exponentially at large $\ell$, similarly to $g_T^{\rm approx}(\ell)$.  

As explained in \cite{Son:2002sd}, singularities of the retarded Green's function in the complex $\omega$ plane correspond to quasi-normal modes of the corresponding field on the gravity side of the AdS/CFT duality.  In the present context, the quasi-normal modes in question are fluctuations which are purely infalling at the worldsheet horizon and behave as $y^3$ for small $y$.  Such fluctuations occur at values of $\omega$ where $C_R(\omega)$ diverges.  Equivalently, in terms of $w=\omega z_H/y_S$, the quasi-normal modes arise at poles of $c_R(w)$.  A preliminary numerical analysis yielded the following approximate values of the first two quasi-normal modes $w_n$:
 \eqn{FirstTwoW}{
  w_1 = 2.620 - 2.302 i \qquad
  w_2 = 4.650 - 4.316 i \,.
 }
For each $w_n$, $-w_n^*$ is also a quasi-normal frequency.  (Quasi-normal modes of the trailing string were also considered in \cite{Herzog:2006gh} in the $v \to 0$ limit, although the boundary conditions were different.)  A standard expectation is that $\ell_1 \equiv 1/\Im w_1 \approx 0.44$ sets a scale for exponential decay of the retarded Green's function.  This is considerably smaller than $\ell_*$; note however that $\ell_1$ relates only to asymptotic statements at large $\ell$, whereas $\ell_*$ is an approximate characterization of the main structures visible in $g_T(\ell)$.  (Also note that because $|\Im w_1| > 2$, the singularity of $g_T(w)$ in the complex $w$ plane closest to the real axis is not $w_1$ but $\pm 2i$ coming from the factor $\coth(\pi w/2)$, so a decay $g_T(\ell) \sim e^{-2|\ell|}$ is expected rather than $e^{-|\ell|/\ell_1}$.)

\section{Discussion}
\label{DISCUSSION}

One should be able to extract the transport coefficients $\kappa_L$, $\kappa_T$, $\hat{q}_T$, and $\hat{q}_L$ from the low-frequency limit of the symmetrized Wightman functions:
 \eqn{ExtractKappa}{
  \kappa_T = {v \over 2} \hat{q}_T = 
    \lim_{\omega \to 0} G_T(\omega) \qquad
  \kappa_L = v \hat{q}_L = \lim_{\omega \to 0} G_L(\omega) \,.
 }
The logic supporting \eno{ExtractKappa} is a treatment of momentum diffusion as in equations \eno{PartonLangevin}-\eno{qhatExtracted}, only with stochastic averages replaced by the symmetrized Wightman functions.  Using \eno{BestFormG} and $\lim_{w\to 0} g_T(w) = 1$, one finds the results quoted in \eno{HugeKappaL}:
 \eqn{HKL}{
  \hat{q}_T = {2\kappa_T \over v} = 
    2\pi \sqrt{g_{YM}^2 N}\, T^3 \sqrt\gamma/v \qquad
  \hat{q}_L = {\kappa_L \over v} = 
    \pi \sqrt{g_{YM}^2 N}\, T^3 \gamma^{5/2}/v \,.
 }
Because of the dramatic failure of the Einstein relation \eno{EinsteinRelation}, it doesn't make sense to plug the trailing string predictions for $\eta_D$, $\kappa_L$, and $\kappa_T$ into a Langevin description of a finite mass quark: it won't equilibrate to a Maxwell-Boltzman distribution, due to the largeness of $\kappa_L$ as compared to $\eta_D$ at highly relativistic speeds.  Perhaps what is needed is a stochastic treatment of the quasi-normal modes of the trailing string, the lowest of whose frequencies were found in \eno{FirstTwoW}.

Because strongly coupled ${\cal N}=4$ gauge theory is different in significant respects from QCD, there is considerable uncertainty in how one should translate results like \eno{StringEta} and~\eno{ExtractKappa} into quantitative predictions for QCD.  To characterize this uncertainty, consider the following two comparison schemes:
 \begin{itemize}
  \item An ``obvious'' comparison scheme is to equate the temperature and the Yang-Mills coupling.  I will assume $T = 250\,{\rm MeV}$ and $\alpha_s = 0.5$, which are in a representative range for RHIC physics.  To summarize:
 \eqn{ObviousScheme}{
  \hbox{``obvious'' scheme:}\qquad
   T_{{\cal N}=4} = T_{\rm QCD} = 250\,{\rm MeV} \qquad
   g_{YM}^2 N = 12\pi \alpha_s = 6\pi \,.
 }
  \item An ``alternative'' comparison scheme was proposed in \cite{Gubser:2006qh}.  The first part of this prescription is to equate energy density instead of temperature.  Thus in any formula (for example \eno{HKL}) in ${\cal N}=4$ into which $T$ enters, one eliminates it in favor of a power of the energy density before comparing to QCD.  This is approximately equivalent to setting $T_{{\cal N}=4} = T_{\rm QCD}/3^{1/4}$.  The second part of the prescription is to determine $g_{YM}^2 N$ by matching string theory results for the force between a static quark and anti-quark to lattice results.  This match is conspicuously imperfect, but at $T_{\rm QCD} = 250\,{\rm MeV}$ a range $3.5 < g_{YM}^2 N < 8$ was found by comparing static potentials at $r \approx 0.25\,{\rm fm}$, with $g_{YM}^2 N = 5.5$ suggested as a typical value.  To summarize:
 \eqn{AlternativeScheme}{
  \hbox{``alternative'' scheme:}\qquad
   3^{1/4} T_{{\cal N}=4} = T_{\rm QCD} = 250\,{\rm MeV} \qquad
    g_{YM}^2 N = 5.5 \,.
 }
 \end{itemize}

Combining \eno{HKL} with either \eno{ObviousScheme} or \eno{AlternativeScheme} leads to the following estimates of $\hat{q}_T$ for a charm quark propagating through a real QGP:
 \eqn{TwoPrescriptions}{\seqalign{\span\TT\qquad & \span\TL & \span\TR &\qquad\span\TT}{
  ``obvious:'' & \hat{q}_T &\approx (2.2\,{\rm GeV}^2/{\rm fm})
    {1 \over v \sqrt[4]{1-v^2}} = 5.9\,{\rm GeV}^2/{\rm fm} &
    for $p_c = 10\,{\rm GeV}/c$  \cr
  ``alternative:'' & \hat{q}_T &\approx (0.51\,{\rm GeV}^2/{\rm fm})
    {1 \over v \sqrt[4]{1-v^2}} = 1.4\,{\rm GeV}^2/{\rm fm} &
    for $p_c = 10\,{\rm GeV}/c$,
 }}
and for $\hat{q}_L$:
 \eqn{TwoPrescriptionsL}{\seqalign{\span\TT\qquad & \span\TL & \span\TR &\qquad\span\TT}{
  ``obvious:'' & \hat{q}_L &\approx (1.1\,{\rm GeV}^2/{\rm fm})
    {1 \over v (1-v^2)^{5/4}} = 150\,{\rm GeV}^2/{\rm fm} &
    for $p_c = 10\,{\rm GeV}/c$  \cr
  ``alternative:'' & \hat{q}_L &\approx (0.26\,{\rm GeV}^2/{\rm fm})
    {1 \over v (1-v^2)^{5/4}} = 36\,{\rm GeV}^2/{\rm fm} &
    for $p_c = 10\,{\rm GeV}/c$,
 }}
where I have set $m_c = 1.4\,{\rm GeV}$.  The choice $p_c = 10\,{\rm GeV}/c$ corresponds roughly to $p_{\rm T} = 5\,{\rm GeV}$ for non-photonic electrons.  The values of $\hat{q}_T$ and $\hat{q}_L$ scale with temperature as $(T_{\rm QCD}/250\,{\rm MeV})^3$ in both the ``obvious'' and ``alternative'' prescriptions.  Based on the large values for $\hat{q}_L$ in \eno{TwoPrescriptionsL}, especially using the ``obvious'' comparison prescription, a conservative conclusion is simply that longitudinal fluctuations in the force are strong and merit a more thorough investigation.

There appears to be some tension between the results of the present work and those of \cite{Liu:2006ug,LiuNew}.  We insert a color source in ${\cal N}=4$ gauge theory in the same way, namely by letting a fundamental string end on the boundary.  This is why our results have the same $\sqrt{g_{YM}^2 N}$ scaling.  But the functional dependence as well as the magnitude of our results differ significantly: in \cite{Liu:2006ug,LiuNew} it was found for partons in the adjoint representation that
 \eqn{LRWqhat}{
  \hat{q}_{\rm LRW} = {\pi^{3/2} \Gamma(3/4) \over \Gamma(5/4)}
    \sqrt{g_{YM}^2 N} \, T^3
 }
in the light-like limit, $v \to 1$.  Combining \eno{LRWqhat} with either \eno{ObviousScheme} or \eno{AlternativeScheme} leads to
 \eqn{LRWqhatNumerical}{\seqalign{\span\TT\qquad & \span\TL & \span\TR}{
  ``obvious:'' & \hat{q}_{\rm LRW} &
    \approx 2.6\,{\rm GeV}^2/{\rm fm}  \cr
  ``alternative:'' & \hat{q}_{\rm LRW} &
    \approx 0.61\,{\rm GeV}^2/{\rm fm} \,.
 }}
Again it is worth noting the sensitive dependence $(T_{\rm QCD}/250\,{\rm MeV})^3$ of $\hat{q}_{\rm LRW}$.  It is also important to recall that $\hat{q}_{\rm LRW}$ is defined differently from $\hat{q}_T$, in terms of a partially light-like Wilson loop rather than momentum broadening.

As a consistency check on the comparison of $\hat{q}_T$ to a transport quantity used in modeling a real-world quark-gluon plasma, one should inquire whether the path lengths of real charm quarks are longer than the typical time scale $t_*$ of force-force correlators.  An estimate of this time scale can be read off from the quantity $\ell_*$ discussed following \eno{gtReal} as $t_* = \ell_* z_H/y_S$.  Combining this with either \eno{ObviousScheme} or \eno{AlternativeScheme} leads to
 \eqn{LabScale}{\seqalign{\span\TT\qquad & \span\TL & \span\TR}{
  ``obvious:'' & t_* &= (1.0\,{\rm fm}/c) \sqrt{E / 10\,{\rm GeV}}
     \cr
  ``alternative:'' & t_* &= (1.3\,{\rm fm}/c)
    \sqrt{E / 10\,{\rm GeV}} \,.
 }}
Clearly these times are short compared to the lifetime of the QGP, which is roughly $6\,{\rm fm}/c$ in central collisions.  But it is more relevant to compare to the mean path length of the charm quarks which escape from the QGP and are detected.  This is presumably bounded below by the corresponding quantity for hard partons without a flavor tag, estimated in \cite{Dainese:2004te} as $\lambda_{\rm mean} = 2.5\,{\rm fm}$.  Perhaps an even more relevant comparison is the relaxation time $t_D = 1/\eta_D$ for charm quarks as calculated via the trailing string picture.  This quantity was estimated in \cite{Gubser:2006qh} as $t_D = 0.6\,{\rm fm}/c$ using the ``obvious'' prescription and $2.1\,{\rm fm}/c$ using the ``alternative'' prescription.  The comparison with $t_D$ is the most stringent: 
 \eqn{tCompare}{\seqalign{\span\TT\qquad & \span\TL & \span\TR}{
  ``obvious:'' & {t_* \over t_D} &= 1.7 \sqrt{E / 10\,{\rm GeV}}
     \cr
  ``alternative:'' & {t_* \over t_D} &= 0.63
    \sqrt{E / 10\,{\rm GeV}} \,.
 }}
The conclusion from \eno{tCompare} is that, depending somewhat on assumptions, time correlations in the fluctuating forces on a charm quark probably matter, especially at higher energies.  It would be useful to inquire what happens when the quark mass is made finite explicitly in the string construction by inserting a flavor brane as in \cite{Karch:2002sh,Herzog:2006gh}.

For $b$ quarks, whose mass is roughly $4.8\,{\rm GeV}$, time correlations are less important for the energies attainable at RHIC: $t_*$ scales as $1/\sqrt{m}$ while $t_D$ scales as $m$, so for fixed energy, the ratio $t_*/t_D$ is decreased by a factor of $(m_b/m_c)^{3/2} \approx 6.3$.

Correlation times may also help clarify the relation with \cite{Liu:2006ug,LiuNew}: $t_*$ diverges as $v \to 1$, so if the path length $\lambda$ is held fixed as this limit is taken, it conflicts strongly with the $\omega \to 0$ limit in \eno{ExtractKappa}.  It is tempting to speculate that somehow cutting off the integral \eno{kappaDef} at finite time would allow one to define a $\hat{q}$-like quantity with a finite $v \to 1$ limit.

A characteristic feature of radiative descriptions of energy loss following \cite{Baier:1996kr,Baier:1996sk,Zakharov:1997uu} is the scaling of $\Delta E$ not as path length $\lambda$ but as $\lambda^2$.  The Brodsky-Hoyer inequality \cite{Brodsky:1992nq}, 
 \eqn{BHineq}{
  \Delta E < {1 \over 2} \hat{q} \lambda^2 \,,
 }
derived from uncertainty principle considerations, is saturated to within a factor of order unity in the BDMPS formalism.  This is in contrast to a Langevin description including drag force, where $\Delta E \propto \lambda$.  As a rough comparison between a radiative description and the trailing string description, it is interesting to ask above what critical path-length $\lambda_c$ the inequality \eno{BHineq} is satisfied if one extracts the left-hand side from the drag force (ignoring longitudinal fluctuations in the force) and sets $\hat{q}=\hat{q}_T$.  The answer is
 \eqn{lambdaC}{
  \lambda_c = {v^2 \sqrt\gamma \over 2T} = {v^2 \over 2Ty_S} \,.
 }
For $p_c = 10\,{\rm GeV}$, one finds $\lambda_c = 1.0\,{\rm fm}$ using the ``obvious'' scheme and $\lambda_c = 1.4\,{\rm fm}$ using the ``alternative'' scheme.  In principle one can distinguish between predictions of radiative and Langevin descriptions of energy loss by making independent measurements of energy loss and momentum broadening.  It has already been suggested \cite{Schweda:2006qc} in the context of a Langevin description that two-point correlators in azimuthal angle for $D$-$\bar{D}$ pairs could provide some insight into the drag force.  Further investigation of the potential uses of heavy quark two-point correlators to distinguish between competing models of energy loss would clearly be of interest.

The trailing string makes the distinctive prediction that, for heavy quarks, both the average drag force and fluctuations are enhanced by powers of the Lorentz factor $1/\sqrt{1-v^2}$ for $v$ close to $1$.  This clearly implies that heavy quarks feel dissipative effects more strongly when they are more energetic.  So the expectation from the trailing string picture is that $R_{AA}$ as a function of $p_{\rm T}$ for non-photonic electrons has a more negative slope than predicted by treatments such as \cite{Armesto:2005mz} in which $\hat{q}$ is held fixed (i.e.~independent of $p_{\rm T}$).  It is therefore gratifying to observe in figure~3 of \cite{Adare:2006nq} a persistently negative slope of $R_{AA}$ over a wide range of momenta, $1.5\,{\rm GeV}/c < p_{\rm T} < 7\,{\rm GeV}/c$.  The predictions of \cite{Armesto:2005mz} include a weaker dependence of $R_{AA}$ on $p_{\rm T}$ over this range.\footnote{In referring to the predictions of one or another theoretical study, I am relying upon the portrayal of those predictions in \cite{Adare:2006nq} or \cite{Abelev:2006db}.  This means, in particular, a choice $\hat{q} = 14\,{\rm GeV}^2/{\rm fm}$ in the calculations of \cite{Armesto:2005mz}.}  The predictions of \cite{Moore:2004tg} are also rather flat in this interval, despite a dependence of $\eta_D(p)$ on $p$ that is only slowly decreasing in the range of interest.  The predictions of \cite{vanHees:2005wb}, which, like \cite{Moore:2004tg}, are based on a Langevin approach, have a significantly larger momentum range in which $R_{AA}$ is strongly decreasing, and $\eta_D(p)$ as presented in the closely related work \cite{vanHees:2004gq} is again slowly decreasing in the range of interest.  Only three points, with $p_{\rm T}$ roughly between $4.8$ and $6.5\,{\rm GeV}/c$, lie below the predictions of \cite{vanHees:2005wb}; the same three, plus perhaps one other at $p_{\rm T} \approx 4.2\,{\rm GeV}/c$, lie below the predictions of \cite{Armesto:2005mz}.  In summary, the comparison of $R_{AA}$ as reported in \cite{Adare:2006nq} with existing phenomenological models reinforces the view that dissipative effects increase with the momentum of the heavy quark; but without some more systematic study, it is hard to tell whether the trailing string picture does a better job of explaining the data than other approaches.

Results from the STAR collaboration \cite{Abelev:2006db} agree fairly well with \cite{Armesto:2005mz} if the contribution from bottom quarks is removed.  But in light of the comparisons of $pp \to eX$ PHENIX data \cite{Adare:2006hc} with fixed-order-plus-next-to-leading-log calculations \cite{Cacciari:2005rk}, it seems ad hoc to neglect bottom contributions completely.  The authors of \cite{Armesto:2005mz} cogently warn of the difficulty of disentangling the contributions of charm and bottom in the absence of vertex reconstruction.  In any case, the STAR data are at least consistent with a more negative slope for $R_{AA}$ as a function of $p_{\rm T}$ than either the $c+b$ or $c$ only curves from \cite{Armesto:2005mz}.

\section{Version history}
\label{VERSIONS}

In the initial version of this work it was claimed incorrectly that $\hat{q}_T$ scales as $1/\sqrt{1-v^2}$ as $v \to 1$.  The correct scaling $1/\sqrt[4]{1-v^2}$ was worked out in private communications with J.~Casalderrey-Solana and D.~Teaney, who independently obtained key aspects of the current work.  Subsequently and independently, they and I worked out the case of longitudinal fluctuations.  The crucial correction to my original treatment of transverse fluctuations is the replacement $\coth {\omega \over 2T} \to \coth {\omega \over 2T y_S}$ in \eno{WightmanFromGR}, as justified in appendix~\ref{SCHWINGER} following \cite{Casalderrey-Solana:2007qw}.  Inspired by their treatment, I realized that the Green's functions could be expressed as functions of $w=\omega z_H/y_S$ up to overall powers of $z_H$ and $y_S$.  This facilitated an analysis of the characteristic time-scale of force correlations.  It also occurred to me that a simple zero-temperature treatment sufficed to understand the leading behavior at large $w$ of $g_T(w)$.  Finally, in private communications, H.~Liu, K.~Rajagopal, and U.~Wiedemann all urged me to distinguish even more carefully between $\hat{q}$ as computed through a Wilson loop formalism and $\hat{q}_T \equiv \langle p_\perp^2 \rangle / \lambda$.  Taken together, these developments seemed an appropriate occasion for the substantial rewrite that occurred between arXiv versions 2 and 3.

\section*{Acknowledgments}

I thank C.~Herzog, H.~Liu, J.~Maldacena, G.~Michalogiorgakis, C.~Nayak, K.~Rajagopal, U.~Wiedemann, and especially J.~Casalderrey-Solana and D.~Teaney for useful discussions.  I am grateful to B.~Zajc for pointing out reference \cite{Schweda:2006qc} to me, for correspondence regarding the electron's typical momentum fraction in $D$-meson decays, and for comments regarding the interesting $p_{\rm T}$ dependence of $R_{AA}$ for non-photonic electrons as reported in \cite{Adare:2006nq}.  I am indebted to J.~Friess, G.~Michalogiorgakis, and S.~Pufu for collaboration on work that led to the results appearing in appendix~\ref{WORLDSHEET}.  This work was supported in part by the Department of Energy under Grant No.\ DE-FG02-91ER40671, and by the Sloan Foundation.

\clearpage
\appendix
\section{Infalling and outgoing coordinates}
\label{WORLDSHEET}

A crucial property of the wave-functions $\Psi_R(\omega,t,y)$ appearing in \eno{SeparateAnsatz} is that they depend only on the infalling coordinate close to the horizon.  The purpose of this appendix is to demonstrate that the infalling and outgoing coordinates are quoted correctly in \eno{InOutCoordinates}, to leading order in small $y_S-y$, where as usual $y_S = \sqrt[4]{1-v^2}$.  Actually, I will do somewhat more by explicitly constructing coordinates in which the worldsheet metric is seen to be locally conformal to ${\bf R}^{1,1}$ (as any two-dimensional metric with $-$$+$ signature must be).

It is convenient to start by parametrizing the worldsheet using $x^1$ and $y$ rather than $t$ and $y$.  The worldsheet metric is easily seen to be
 \eqn{dsInduced}{
  ds_2^2 = g_{\alpha\beta} d\sigma^\alpha d\sigma^\beta
   = {L^2 \over z_H^2 y^2} \left[ \left( 1 - {h \over v^2}
    \right) (dx^1)^2 - {2 z_H y^2 \over v} dx^1 dy + z_H^2 dy^2
   \right] \,,
 }
where as usual $h = 1-y^4$.  The manipulations to cast the metric \eno{dsInduced} in a conformally flat form are relatively simple because $g_{\alpha\beta}$ depends on $y$ but not $x^1$:
 \eqn{IMCS}{
  ds_2^2 &= {L^2 \over y^2} \left( 1 - 
     {y^4/v^2 \over 1-h/v^2} \right) dy^2 + 
    {L^2 \over z_H^2 y^2} \left( 1 - {h \over v^2} \right)
     \left( dx^1 - {z_H y^2 / v \over 1-h/v^2} dy \right)^2
       \cr
   &= {L^2 \over y^2} {1-v^2 \over h-v^2} dy^2 - 
     {L^2 \over v^2 z_H^2 y^2} (h-v^2) dq^2  \cr
   &= \Omega^2 \left( -dq^2 + d\eta^2 \right)
 }
where 
 \eqn{WimcsDef}{
  \Omega^2 = {L^2 \over y^2} {h-v^2 \over v^2 z_H^2}
 }
and I have defined new coordinates
 \eqn{qDef}{
  q = x^1 + \int dy {z_H y^2 v \over h-v^2} \qquad
  \eta = \pm v z_H \sqrt{1-v^2} \int {dy \over h-v^2} \,.
 }
Because $h-v^2=y_S^4-y^4$, the integrals in \eno{qDef} diverge at the horizon.  So one must define $q$ and $\eta$ piecewise in two regions:
 \eqn{RegionIDefs}{\left. \eqalign{
  q_{\bf I} &= x^1 + {z_H v \over 4 i y_S} \left( \log {1-iy/y_S \over 
    1+iy/y_S} + i \log {1+y/y_S \over 1-y/y_S} \right)  \cr
  \eta_{\bf I} &= {z_H v \over 4 i y_S} \left( -\log {1-iy/y_S \over 
    1+iy/y_S} + i \log {1+y/y_S \over 1-y/y_S} \right)} \right\}
   \qquad\hbox{for $0 < y < y_S$;}
 }
 \eqn{RegionIIDefs}{\left. \eqalign{
  q_{\bf II} &= x^1 + 
   {z_H v \over 4 i y_S} \left( \log {1-iy/y_S \over 
    1+iy/y_S} + i \log {1+y/y_S \over -1+y/y_S} \right)  \cr
  \eta_{\bf II} &= {z_H v \over 4 i y_S} \left( \log {1-iy/y_S \over 
    1+iy/y_S} - i \log {1+y/y_S \over -1+y/y_S} \right)} \right\}
   \qquad\hbox{for $y_S < y < 1$,}
 }
The second form of the metric in \eno{IMCS} is analogous to the standard form of the Schwarzschild metric.  In region ${\bf I}$ (outside the horizon) $y$ is the spacelike variable and $q$ is the timelike variable.  In region ${\bf II}$ (inside the horizon) $y$ is timelike and $q$ is spacelike.

Using $x^1=vt+\xi(y)$, one may extract expressions for $q_{\bf I}$ in terms of $t$ and $y$ rather than $x^1$ and $y$; $\eta_{\bf I}$, on the other hand, is a function only of $y$.  To construct infalling and outgoing coordinates in region ${\bf I}$, one simply forms
 \eqn{qPM}{
  q_\pm = q_{\bf I} \pm \eta_{\bf I} \,.
 }
Expanding in small $y_S-y$, one finds
 \eqn{quPM}{
  q_- = vt + \ldots \qquad
  q_+ = v \left[ t - {z_H \over 2y_S} \log(y_S-y) \right] + 
    \ldots \,,
 }
where the omitted terms comprise constant terms (i.e.~independent of $t$ and $y$) and terms involving positive powers of $y_S-y$.  Up to the overall factor of $v$, the expressions in \eno{quPM} are precisely the ones given in \eno{InOutCoordinates}.

\section{Relations between real-time thermal propagators}
\label{SCHWINGER}

The usual relations among the Green's functions defined in \eno{SeveralGreen} may be summarized as follows:\footnote{I am indebted to J.~Casalderrey-Solana and D.~Teaney for pointing out that the second equality in equation (25) of the original version of this paper was in error, and for private communications related to the contents of this appendix.  The treatment in this appendix follows in its essential points that of \cite{Casalderrey-Solana:2007qw}.}
 \eqn[c]{Grels}{
  G^F(\omega) = \Re G^R(\omega) + i \coth {\omega \over 2T}
    \Im G^R(\omega)  \cr
  G(\omega) = -\Im G^F(\omega) = 
    -\coth {\omega \over 2T} \Im G^R(\omega) \,.
 }
The first equality in \eno{Grels} usually can be derived by inserting complete sets of states; the second is immediate from the definitions; and the third is an obvious corollary of the first.  However, the usual derivation of the first equality relies crucially on having a fully equilibrated system, where $\langle Q \rangle = \tr(e^{-\beta H} Q) / \tr e^{-\beta H}$ for any operator $Q$, and the heavy quark is far from equilibrated.  It turns out that a very simple modification of \eno{Grels} suffices: one must make the replacement $T \to T y_S$ in \eno{Grels}, where $y_S = \sqrt[4]{1-v^2}$.  This applies equally for transverse and longitudinal fluctuations, and I will not need to distinguish between them in this appendix.

\begin{figure}
  \centerline{\includegraphics[width=7in]{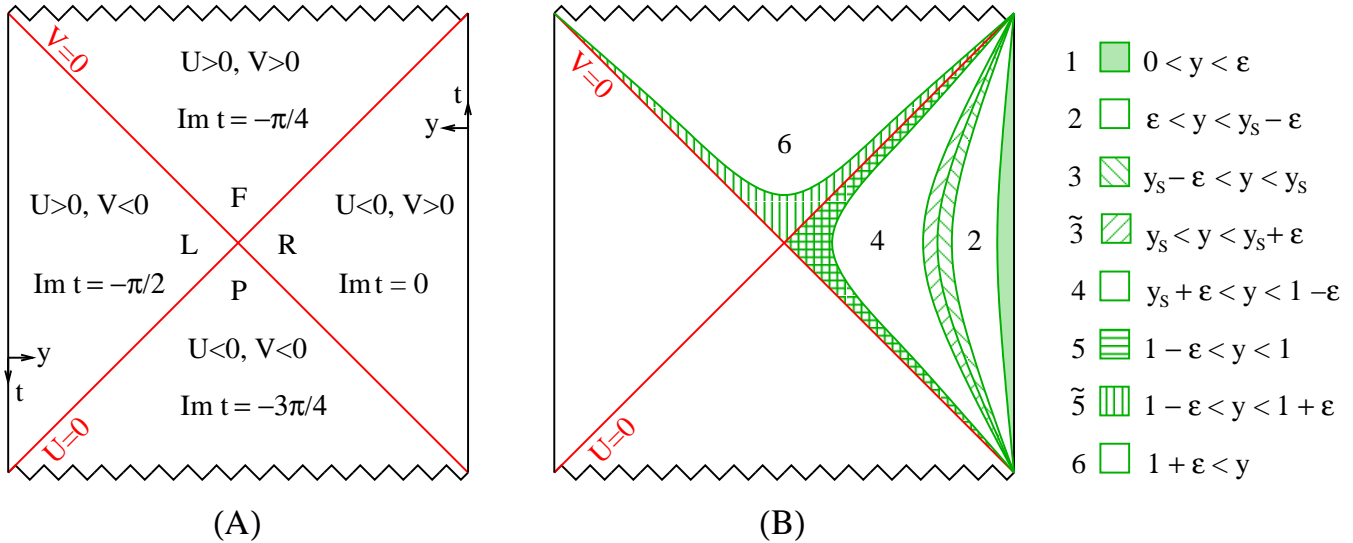}}
  \caption{(A) The Penrose diagram for $AdS_5$-Schwarzschild with conventions on the phase of $t$ specified.  Increasing the real part of $t$ corresponds to moving upward on the R boundary and downward on the L boundary.  Increasing $y$ means moving to the left in R and to the right in L.  (B) The odd-numbered regions are narrow slices of the Penrose diagram corresponding to series expansions around $y=0$, $y_S$, and $1$.  The even-numbered regions fill in between these series expansions.  Note that $3$ and $\tilde{3}$ are disjoint, but $5$ is a subset of $\tilde{5}$.  The corresponding regions $-1$ to $-6$ are the reflections of $1$ to $6$ through the point $U=V=0$.}\label{KruskalCoords}
 \end{figure}

A derivation of the modified form of \eno{Grels} may be carried out using the Schwinger-Keldysh formalism, as adapted to AdS/CFT in \cite{Herzog:2002pc} following earlier work \cite{Horowitz:1998xk,Maldacena:2001kr,Balasubramanian:1998de}.  For simplicity, let's set $z_H=1$, so $T = 1/\pi$.  First define Kruskal coordinates $U$ and $V$ such that
 \eqn{KruskalDef}{
  UV = {y-1 \over y+1} e^{-2\tan^{-1} y} \qquad
  {V \over U} = -e^{4t} \,.
 }
To make the most efficient use of these relations, it helps to note the following points:
 \begin{enumerate}
  \item One may extend the second relation in \eno{KruskalDef} over the entire Penrose diagram by assigning $t$ an imaginary part which is piecewise constant but changes by an odd-integer multiple of $i\pi/4$ as $U$ or $V$ switches sign.  A consistent way of doing so is exhibited in figure~\ref{KruskalCoords}a.
  \item The first relation in \eno{KruskalDef} defines a monotonically increasing map from $UV \in (-1,e^{-\pi})$ to $y \in (0,\infty)$, and $UV=0$ maps to $y=1$.
  \item As a result of the previous two points, the relations \eno{KruskalDef} define a one-to-one map $(U,V) \to (t,y)$, where the domain of $(U,V)$ is points in ${\bf R}^2$ subject only to the restriction $-1< UV < e^{-\pi}$, and the range of $(t,y)$ has four disjoint pieces corresponding to the four regions (RFLP) in figure~\ref{KruskalCoords}a.  The R piece, for example, comprises points in ${\bf R}^2$ subject only to the restriction $0<y<1$.
 \end{enumerate}

The trailing string solution \eno{TrailingString} may be recast in the form
 \eqn{CleverTrailingString}{
  x^1 = {v \over 2} \log V + v \tan^{-1} y \,.
 }
(Recall that $z_H=1$.)  In this form it is clear that the trailing string is entirely non-singular at the horizon between the R and F regions.  The singularity at $V=0$, corresponding to $\Re t \to -\infty$, is genuine.  In the spirit of the Schwinger-Keldysh formalism, one prepares a state of the system at $\Re t = -\infty$ and propagates it first forward along $\Im t = 0$ and then back along $\Im t = -\pi/2$.  Correspondingly, one may use \eno{CleverTrailingString} in the R and F regions and
 \eqn{LPstring}{
  x^1 = {v \over 2} \log(-V) + v \tan^{-1} y - iv\pi/2
 }
in the L and P regions.  The two solutions \eno{CleverTrailingString} and \eno{LPstring} can be matched at $V=0$ by continuing $V$ through the lower half-plane---that is, for $\Im V$ slightly negative.  Such a continuation is consistent with the conventions laid out in figure~\ref{KruskalCoords} because in the L region one may rewrite \eno{LPstring} in terms of $t$ and $y$ in precisely the original form \eno{TrailingString}.

Let's now consider perturbation wave-functions $\Psi_\alpha$ for oscillations of the string in a slightly more general fashion than in section~\ref{FLUCTUATE}.  The index $\alpha$ will run over values in a rather large set:
 \eqn{alphaSet}{
  \alpha \in \{ \pm 1, \pm 2, \pm 3, \pm \tilde{3}, \pm 4,
   \pm 5, \pm \tilde{5}, \pm 6 \} \,.
 }
These values correspond to different regions of the Penrose diagram, as indicated in figure~\ref{KruskalCoords}b.  In each of the odd-numbered regions, which are all narrow slices near some definite value of $y$, one may expand the general solution to the wave equation for $\Psi$ to leading order as follows:
 \eqn{SeveralExpansions}{\seqalign{\span\TL & \span\TR\qquad & \span\TL & \span\TR}{
  \Psi_1 &= [A_1 + B_1 y^3] e^{-i\omega t} &
   \Psi_{-1} &= [A_{-1} + B_{-1} y^3] e^{-i\omega t}  \cr
  \Psi_3 &= [A_3 + B_3 (y_S-y)^{i\omega/2y_S}] e^{-i\omega t} &
   \Psi_{-3} &= [A_{-3} + B_{-3} (y_S-y)^{i\omega/2y_S}]
     e^{-i\omega t}  \cr
  \Psi_{\tilde{3}} &= [A_{\tilde{3}} + B_{\tilde{3}}
    (y-y_S)^{i\omega/2y_S}] e^{-i\omega t} &
  \Psi_{-\tilde{3}} &= [A_{-\tilde{3}} + B_{-\tilde{3}}
    (y-y_S)^{i\omega/2y_S}] e^{-i\omega t}  \cr
  \Psi_5 &= [A_5 + B_5 (1-y)] (1-y)^{-i\omega/4} e^{-i\omega t} &
   \Psi_{-5} &= [A_{-5} + B_{-5} (1-y)] (1-y)^{-i\omega/4}
    e^{-i\omega t}  \cr
  \Psi_{\tilde{5}} &= [A_{\tilde{5}} + B_{\tilde{5}} (1-y)] 
     V^{-i\omega/2} &
   \Psi_{-\tilde{5}} &= [A_{-\tilde{5}} + B_{-\tilde{5}} (1-y)] 
     (-V)^{-i\omega/2} \,.
 }}
For brevity I have dropped subleading terms from the $A$-type solutions in \eno{SeveralExpansions} even when they are bigger than the $B$-type solutions.  There is no loss of generality in assuming $e^{-i\omega t}$ dependence because the entire spacetime has a timelike Killing vector which is $\partial/\partial t$ in each patch.

The expressions in \eno{SeveralExpansions} involve imaginary powers of quantities which are real and positive.  Such expressions are unambiguously defined: $x^{i\alpha} \equiv e^{i\alpha\log x}$.  To specify $\Psi$ globally one must prescribe via some analytic continuation how the pairs $(\Psi_3,\Psi_{\tilde{3}})$, $(\Psi_{-3},\Psi_{-\tilde{3}})$, and $(\Psi_{\tilde{5}},\Psi_{-\tilde{5}})$ fit together.  Also one must solve the wave equation for $\Psi$ in the intermediate regions $\pm 2$, $\pm 4$, either through some approximation or via numerics.\footnote{One could also find numerical solutions in region $\pm 6$ and perhaps match to some expansion near the future and past spacelike singularities.  This would perhaps be of interest in an exploration of the physics of the black hole singularity, and it could also be of some relevance in computing higher point functions of operators pertaining to the heavy quark, but such matters are beyond the scope of the current work.}  The result of such prescriptions and approximate solutions can be cast in terms of $2 \times 2$ matrices:
 \eqn{TransferMatrices}{
  \vec{v}_\alpha = {\bf M}_{\alpha\beta}\, \vec{v}_\beta 
    \qquad\hbox{where}\quad
  \vec{v}_\alpha = 
    \begin{pmatrix} A_\alpha \\ B_\alpha \end{pmatrix} \,.
 }
Here $\alpha$ and $\beta$ are in the set \eno{alphaSet}, excluding the even numbered values.  Note that ${\bf M}_{\alpha\beta}$ for each $\alpha$ and $\beta$ is a $2 \times 2$ matrix: $\alpha$ and $\beta$ label the matrix, not its components.  An obvious property of the ${\bf M}_{\alpha\beta}$ is ${\bf M}_{\beta\alpha} = {\bf M}_{\alpha\beta}^{-1}$.

The properties of the matrices ${\bf M}_{\alpha\beta}$ that I will require are:
 \eqn{MinusM}{
  {\bf M}_{-1,-3} = {\bf M}_{13} \qquad
   {\bf M}_{-\tilde{3},-5} = {\bf M}_{\tilde{3}5}
 }
 \eqn{HorizonPassing}{
  {\bf M}_{3\tilde{3}} = 
   \begin{pmatrix} 1 & 0 \\ 0 & e^{-\pi\omega/2y_S} \end{pmatrix}
    \qquad
  {\bf M}_{-3,-\tilde{3}} = 
   \begin{pmatrix} 1 & 0 \\ 0 & e^{\pi\omega/2y_S} \end{pmatrix}
 }
 \eqn{ReparamFive}{
  {\bf M}_{5\tilde{5}} = \left( {e^{-\pi/2} \over 2} 
    \right)^{-i\omega/4} \begin{pmatrix} 1 & 0 \\ 0 & 1 
     \end{pmatrix} \qquad
  {\bf M}_{-5,-\tilde{5}} = \left( {e^{-\pi/2} \over 2}
    \right)^{-i\omega/4} e^{\pi\omega/2}
     \begin{pmatrix} 1 & 0 \\ 0 & 1 \end{pmatrix}
 }
 \eqn{VzeroMatch}{
  {\bf M}_{\tilde{5},-\tilde{5}} = e^{\pi\omega/2}
    \begin{pmatrix} 1 & 0 \\ 0 & 1 \end{pmatrix}
 }
 \eqn{MoneThree}{
  {\bf M}_{13}\, {\bf D} = \begin{pmatrix} 1 & 1 \\
    C_R(\omega) & C_R(\omega)^*-iX \end{pmatrix} \equiv
    \widetilde{\bf M}_{13} \,,
 } 
where ${\bf D}$ is some diagonal matrix, $C_R(\omega)$ is the same quantity as introduced in \eno{FarSoln}, and $X = 2v^2\omega z_H/3(1-v^2)$ as in \eno{XYdef}.  I will justify \eno{MinusM}-\eno{MoneThree} after showing how to obtain the first equation in \eno{Grels} from them.

To calculate Schwinger-Keldysh propagators $G_{\alpha\beta}(\omega)$, with $\alpha$ and $\beta$ equal to $\pm 1$, one should specify $A_1(\omega)$ and $A_{-1}(\omega)$ and calculate the ``response'' functions, $B_1(\omega)$ and $B_{-1}(\omega)$.  To this end one requires the matrix
 \eqn{Moneone}{
  {\bf M}_{-1,1} &= {\bf M}_{-1,-3}\, {\bf M}_{-3,-\tilde{3}}\,
    {\bf M}_{-\tilde{3},-5}\, {\bf M}_{-5,-\tilde{5}}\,
    {\bf M}_{-\tilde{5},\tilde{5}}\, {\bf M}_{\tilde{5}5}\,
    {\bf M}_{5\tilde{3}}\, {\bf M}_{\tilde{3}3}\, {\bf M}_{31}  \cr
   &= {\bf M}_{13}\, ({\bf M}^{-1}_{3\tilde{3}})^2\, {\bf M}_{31}
    = \widetilde{\bf M}_{13} \begin{pmatrix} 1 & 0 \\
       0 & e^{\pi\omega/y_S} \end{pmatrix}
      \widetilde{\bf M}_{13}^{-1} \,.
 }
The second equality in \eno{Moneone} uses \eno{MinusM}, \eno{HorizonPassing}, \eno{ReparamFive}, and \eno{VzeroMatch}; the third equality also uses \eno{MoneThree} and the fact that ${\bf D}$ commutes with the diagonal matrix $M_{3\tilde{3}}$.  Using the relation $\vec{v}_{-1} = {\bf M}_{-1,1} \vec{v}_1$ and the prescribed values of $A_{\pm 1}$, one may extract $B_{\pm 1}$.  For example, if $A_{-1}=0$, then
 \eqn{GotBone}{
  B_1(\omega) = C_F(\omega) A_1(\omega) \qquad\hbox{where}\quad
   C_F(\omega) = {C_R(\omega) e^{\pi\omega/y_S} - 
     C_R(\omega)^* + iX \over e^{\pi\omega/y_S} - 1} \,.
 }
A rephrasing of the result \eno{GotBone} is that the wave-function $\Psi_F(\omega,t,y)$ suitable for computing the time-ordered propagator $G^F(\omega) = G_{11}(\omega)$ has leading behavior
 \eqn{PsiFbehave}{
  G^F(\omega) = [1 + C_F(\omega) y^3] e^{-i\omega t}
 }
near the boundary of the $R$ region.  Then, in precise analogy to \eno{GRextract},
 \eqn{GFsimple}{
  G^F(\omega) = -{\cal G}^{y\beta} \partial_\beta 
    \log \Psi_F(\omega,t,y) \Big|_{y=0} = 
   -Y \left( iX \coth {\pi\omega \over 2y_S} + 
     2 {C_R(\omega) e^{\pi\omega/y_S} - C_R(\omega)^*
       \over e^{\pi\omega/y_S} - 1} \right) \,,
 }
where $Y$ is defined as in \eno{XYdef}.  Comparing with \eno{GotG}, one sees immediately that
 \eqn{GFversusGR}{
  G^F(\omega) = \Re G^R(\omega) + i \coth {\pi\omega \over 2y_S}
    \Im G^R(\omega) \,.
 }
Restoring factors of $z_H$ amounts to replacing $\pi$ by $1/T$ in \eno{GFversusGR}, so this is the desired modification of the first equation in \eno{Grels}.

Now let's return to the justification of the formulas \eno{MinusM}-\eno{MoneThree}.  The main issue is continuing from $\Psi_{\pm 3}$ to $\Psi_{\pm \tilde{3}}$ and from $\Psi_{\tilde{5}}$ to $\Psi_{-\tilde{5}}$.  An argument was made in \cite{Herzog:2002pc}, and extended to the present context in \cite{Casalderrey-Solana:2007qw}, that the correct way to do this is to continue in the lower half of the complex $V$ plane and the upper half of the complex $U$ plane when horizon singularities are encountered.  By so doing one imposes infalling conditions on positive frequency modes and outgoing conditions on negative frequency modes.  Formally one may simply send $U \to U+i\delta$ and $V \to V-i\delta$ for small positive $\delta$.  In R this corresponds to sending $y \to y+i\delta$, and in L it corresponds to sending $y \to y-i\delta$.  Consider the consequences for matching the oscillatory solution from $3$ to $\tilde{3}$, which takes place wholly within the R region:
 \eqn{BthreeMatch}{
  \lim_{y\to y_S-} B_3 (y_S-y-i\delta)^{i\omega/2y_S} 
    = \lim_{y\to y_S+} B_{\tilde{3}} 
       (y-y_S+i\delta)^{i\omega/2y_S} \,,
 }
where the first limit is from below and the second is from above.  From \eno{BthreeMatch} one obtains $B_3 = e^{-\pi\omega/2y_S} B_{\tilde{3}}$, whence the lower right entry in ${\bf M}_{3\tilde{3}}$.  The opposite behavior arises in ${\bf M}_{-3,-\tilde{3}}$ because $y \to y-i\delta$ in the L region.  In crossing the $V=0$ line, one finds $(V-i\delta)^{-i\omega/2} = e^{-\pi\omega/2} (-V+i\delta)^{-i\omega/2}$.  This completes a justification of \eno{HorizonPassing} and~\eno{VzeroMatch}.  The other steps can be justified more straightforwardly, as follows:
 \begin{itemize}
  \item \eno{MinusM} is a consequence of the wave equation for $\Psi$ being the same in regions $\pm 2$ and $\pm 4$.  The non-zero imaginary part of $t$ doesn't affect the equation of motion.
  \item \eno{ReparamFive} is a consequence of applying the rule $x^{i\alpha} = e^{i\alpha\log x}$ only when $x$ is positive.  The details are somewhat intricate, so I will trace the steps for the $A$-type solution from $-5$ to $-\tilde{5}$:
 \eqn{DetailFiveFive}{
  A_{-5} (1-y)^{-i\omega/4} e^{-i\omega t}
    &= A_{-5} (1-y)^{-i\omega/4} e^{-i\omega\Re t} e^{-\pi\omega/2}
     \cr
    &= A_{-5} (-V)^{-i\omega/2} \left( {e^{-\pi/2} \over 2}
       \right)^{i\omega/4} e^{-\pi\omega/2}
    = A_{-\tilde{5}} (-V)^{-i\omega/2} \,,
 }
where in the second equality I used the leading behavior of $V$ near $y=1$ in the L region:
 \eqn{LregionV}{
  (-V)^2 = (1-y) e^{4t} {e^{-\pi/2} \over 2} = 
    (1-y) e^{4\Re t} {e^{-\pi/2} \over 2} \,.
 }
The first and last expressions in \eno{LregionV} are in a form suitable to be raised to the imaginary power $-i\omega/4$.
 \item \eno{MoneThree} is a consequence of the behavior of purely infalling and purely outgoing solutions of the wave equation for $\Psi$.  An infalling solution is one with $B_3=0$, and its behavior the R boundary, \eno{FarSoln}, corresponds to $\vec{v}_1$ given by the first column of the matrix $\widetilde{\bf M}_{13}$.  An outgoing solution should have behavior at the R boundary corresponding to $\vec{v}_1$ given by the second column of $\widetilde{\bf M}_{13}$.  It so happens that if $\Psi_R(\omega,t,y) = e^{-i\omega t} \psi_R(\omega,y)$ is the infalling wave-function constructed in section~\ref{FLUCTUATE}, then
 \eqn{IntertwinedSoln}{
  \Psi_A(\omega,t,y) = e^{-i\omega t} f(y)^{i\omega/2}
    f(y/y_S)^{-i\omega/2y_S} \psi_R(\omega,y)^*
 }
is a purely outgoing solution, where
 \eqn{fDef}{
  f(y) \equiv {1+y \over 1-y} e^{-2 \tan^{-1} y} \,.
 }
Expanding \eno{IntertwinedSoln} for small $y$ gives
 \eqn{SamllYIS}{
  \Psi_A(\omega,t,y) = [1 + (C_R(\omega)^*-iX) y^3] 
    e^{-i\omega t} \,,
 }
as desired. 
 \end{itemize}

\section{Speed limits on single quarks}
\label{TOOFAST}

The string theory calculations presented in this paper pertain to external quarks: pointlike objects in the fundamental representation which have infinite mass associated with their near-field Coulombic color-electric field.  Comparisons with QCD are thus better justified for $c$ and $b$ quarks, whose mass is well above the typical temperature of the QGP, than for light quarks.

If the velocity of an external quark is too close to unity, there is a new reason to be wary of trailing string computations: the horizon on the worldsheet is very close to the boundary, where in a holographic representation of real QCD (assuming there is one) space presumably becomes highly curved, and the simplest calculations based on the Nambu-Goto action in $AdS_5$-Schwarzschild may experience significant corrections.  Let's try to estimate when problems of this sort might start to arise.  It was suggested already in \cite{Maldacena:1997re} to associate an energy scale
 \eqn{muScale}{
  \mu = {L^2/z_H \over 2\pi\alpha'} {1 \over y}
    = {T \over 2} {\sqrt{g_{YM}^2 N} \over y}
 }
with a radius $y$ in $AdS_5$-Schwarzschild.  This is justified by observing that a static string dangling from $y$ into the horizon has mass
 \eqn{mStatic}{
  m_{\rm static} = {L^2/z_H \over 2\pi\alpha'} 
     \left( {1 \over y} - 1 \right) \approx \mu \qquad
   \hbox{for $y \ll 1$.}
 }
Now suppose one identifies a scale $\mu_{\rm fail}$ at which AdS/CFT techniques (at least those based on supergravity and classical strings) start to fail.  From \eno{muScale} one extracts a corresponding $y_{\rm fail}$, and if the worldsheet horizon has $y_S<y_{\rm fail}$ there may be significant corrections to the trailing string results.  Setting $y_S = y_{\rm fail}$ and using \eno{ySdef}, one finds that the Lorentz factor of the heavy quark that the trailing string purports to describe is
 \eqn{gammaFail}{
  \gamma_{\rm fail} = {4 \over g_{YM}^2 N} 
    \left( {\mu_{\rm fail} \over T} \right)^2 \,.
 }
The trouble with \eno{gammaFail} is that $g_{YM}^2 N$, $\mu_{\rm fail}$ and $T$ all incorporate considerable uncertainties.

To get an idea of the range of plausible values for $\gamma_{\rm fail}$, let's consider the two prescriptions \eno{ObviousScheme} and \eno{AlternativeScheme}.  I don't have a systematic way of determining $\mu_{\rm fail}$, but perhaps a reasonable range to consider is $\mu_{\rm fail} = 0.8-1.6\,{\rm GeV}$.  Taking $\mu_{\rm fail} = 1.2\,{\rm GeV}$ as a representative value, one finds
 \eqn{gfEstimates}{\seqalign{\span\TT\qquad & \span\TL & \span\TR}{
  ``obvious:'' & \gamma_{\rm fail} &= 
    4.9 \left( {\mu_{\rm fail} \over 1.2\,{\rm GeV}} \right)^2  \cr
  ``alternative:'' & \gamma_{\rm fail} &=
    29 \left( {\mu_{\rm fail} \over 1.2\,{\rm GeV}} \right)^2 \,.
 }}
Because of the quadratic dependence on both $\mu_{\rm fail}$ and $T$ (not to mention the choice of comparison scheme) $\gamma_{\rm fail}$ remains substantially uncertain.  Evidently, if the ``obvious'' prescription is used, there is some doubt cast on stringy predictions for charm quark when $p_c \gsim 6.7\,{\rm GeV}/c$, corresponding to $p_{\rm T} \gsim 3.4\,{\rm GeV}$.  In the ``alternative'' scheme, there is less reason to worry about stringy corrections near the upper end of the trailing string.

Another thing can go wrong if one wants to represent a finite mass quark as a string ending on a D7-brane, as in \cite{Karch:2002sh,Herzog:2006gh}.  Given $m_c$ for the charm quark, one can use \eno{mStatic} to obtain a corresponding position 
 \eqn{ycDef}{
  y_c = {1 \over 1 + 2m_c/T\sqrt{g_{YM}^2 N}}
 }
of the D7-brane.  If $y_S < y_c$, then there is no horizon on the trailing string: the boundary of the worldsheet is spacelike.  I regard this as a pathology which probably invalidates the trailing string picture for Lorentz factors $\gamma > \gamma_c \equiv 1/y_c^2$.  Estimates of $\gamma_c$ suffer from the same ambiguities as $\gamma_{\rm fail}$, as discussed above.  Using $m_c=1.4\,{\rm GeV}$ and either \eno{ObviousScheme} or~\eno{AlternativeScheme}, one arrives at the estimates
 \eqn{gcEstimates}{\seqalign{\span\TT\qquad & \span\TL & \span\TR}{
  ``obvious:'' & \gamma_c &= 13  \cr
  ``alternative:'' & \gamma_c &= 53 \,.
 }}
These values should again be regarded as incorporating considerable uncertainties: for example, it is puzzling that the horizon causes the  mass of a quark to {\it decrease} from its zero-temperature value, but this decrease is what makes the values in \eno{gcEstimates} higher than those derived from \eno{gfEstimates} with $\mu_{\rm fail} \to m_c$.  The main lesson to draw from \eno{gcEstimates} is that charm's mass is high enough to avoid threatening the existence of the worldsheet horizon for the momenta accessible at RHIC; but attempts to treat light quarks in terms of the trailing string construction are perilous indeed.\footnote{For example, in the ``alternative'' comparison scheme, an up quark whose mass in the medium is assumed to be $m=300\,{\rm MeV}$ would be described in terms of a D7-brane with $y_u = 0.43$, so $\gamma_u = 5.5$ is the maximum Lorentz factor.  It is hardly fair to ignore corrections to relativistic dispersion relations in this context, but if one does so, the result is a total energy of $1.7\,{\rm GeV}$.  As an upper limit on allowed energies for partons described by trailing strings, this is fairly anemic.}

\clearpage
\bibliographystyle{ssg}
\bibliography{qhat}

\end{document}